\documentstyle[12pt]{article}

\makeatletter

\newif\ifnoncomplete

\def\final{\noncompletefalse\typeout{** FINAL form (substyle:Sachiko)}}
\noncompletefalse

\newif\ifrefphysrev

\refphysrevfalse

\def \vol(#1,#2,#3){\ifrefphysrev{{\bf {#1}},
{#3} (19{#2})}\else{{{\bf {#1}}(19{#2}){#3}}}\fi}
\def \NP(#1,#2,#3){Nucl.\ Phys.\          \vol(#1,#2,#3)}
\def \PL(#1,#2,#3){Phys.\ Lett.\          \vol(#1,#2,#3)}
\def \PRL(#1,#2,#3){Phys.\ Rev.\ Lett.\   \vol(#1,#2,#3)}
\def \PRp(#1,#2,#3){Phys.\ Rep.\          \vol(#1,#2,#3)}
\def \PR(#1,#2,#3){Phys.\ Rev.\           \vol(#1,#2,#3)}
\def \PTP(#1,#2,#3){Prog.\ Theor.\ Phys.\ \vol(#1,#2,#3)}
\def \ibid(#1,#2,#3){{\it ibid.}\         \vol(#1,#2,#3)}

\def\thebibliography#1{
\section*{References\@mkboth
  {REFERENCES}{REFERENCES}}\list
  {[\arabic{enumi}]}{\setlength\labelwidth{2ex}
   \setlength\labelsep{0.05in} 
   \setlength\leftmargin{0.25in}
    \setlength\itemsep{0pt}
    \setlength\parsep{0pt}
   \itemsep\parskip
    \usecounter{enumi}}
    \def\newblock{\hskip .11em plus .33em minus -.22em}
    \sloppy
    \sfcode`\.=1000\relax}

\def\@bibitem#1{\item\if@filesw \immediate\write\@auxout
       {\string\bibcite{#1}{\the\c@enumi}}\fi\ignorespaces
       {\ifnoncomplete\reversemarginpar{\hspace*{-1.05in}\makebox[1in][l]
       {{\footnotesize{\sl [#1]}}}}\fi}%
       }

\def\@cite#1#2{\unskip\nobreak\relax
    {[#1]}} 

\def\citenum#1{{\def\@cite##1##2{##1}\cite{#1}}}
\def\citea#1{\@cite{#1}{}}

\newcount\@tempcntc
\def\@citex[#1]#2{\if@filesw\immediate\write\@auxout{\string\citation{#2}}\fi
  \@tempcnta\z@\@tempcntb\m@ne\def\@citea{}\@cite{\@for\@citeb:=#2\do
    {\@ifundefined
       {b@\@citeb}{\@citeo\@tempcntb\m@ne\@citea\def\@citea{,}{\bf ?}\@warning
       {Citation `\@citeb' on page \thepage \space undefined}}%
    {\setbox\z@\hbox{\global\@tempcntc0\csname b@\@citeb\endcsname\relax}%
     \ifnum\@tempcntc=\z@ \@citeo\@tempcntb\m@ne
       \@citea\def\@citea{,}\hbox{\csname b@\@citeb\endcsname}%
     \else
      \advance\@tempcntb\@ne
      \ifnum\@tempcntb=\@tempcntc
      \else\advance\@tempcntb\m@ne\@citeo
      \@tempcnta\@tempcntc\@tempcntb\@tempcntc\fi\fi}}\@citeo}{#1}}
\def\@citeo{\ifnum\@tempcnta>\@tempcntb\else\@citea\def\@citea{,}%
  \ifnum\@tempcnta=\@tempcntb\the\@tempcnta\else
   {\advance\@tempcnta\@ne\ifnum\@tempcnta=\@tempcntb \else \def\@citea{--}\fi
    \advance\@tempcnta\m@ne\the\@tempcnta\@citea\the\@tempcntb}\fi\fi}

\def\affiliation#1{\cr
\makebox[0in]{\parbox{8in}{\begin{center} {\sl #1}\end{center}}} \cr}
\def\@affiliation{}

\def\and{\cr \makebox[0in]{\rule[-1cm]{0mm}{1cm}and } \cr}

\def\maketitle{\par
 \begingroup
 \def\thefootnote{\fnsymbol{footnote}}
 \def\@makefnmark{\hbox
 to 0pt{$^{\@thefnmark}$\hss}}
 \if@twocolumn
 \twocolumn[\@maketitle]
 \else \newpage
 \global\@topnum\z@ \@maketitle \fi\thispagestyle{plain}\@thanks
 \endgroup
 \setcounter{footnote}{0}
 \let\maketitle\relax
 \let\@maketitle\relax
 \gdef\@thanks{}\gdef\@author{}\gdef\@title{}
 \gdef\@affiliation{} \let\affiliation\relax	%
 \let\thanks\relax}

\def\@maketitle{\newpage
 \null
 \vskip 0em plus 2em minus 0em     
 \ifx\@date\@empty\else
   \begin{flushright}
    {\ifnoncomplete(\today)
     \else{{\normalsize \@date}\\}\fi}      
   \end{flushright}
   \vskip 3em plus 2em minus 2em   
 \fi
 \begin{center}
  {\frtnsfb \@title \par}     
  \vskip 3em plus 1em minus 1.5em  
  {
   \lineskip .5em plus 0em minus .3em   
   \begin{tabular}[t]{c}\@author
   \end{tabular}\par}
\end{center}
 \par
 \vskip 6em plus 2em minus 4em}     

\def\abstract{\if@twocolumn
\section*{Abstract}
\else \normalsize
\fi}

\def\endabstract{\if@twocolumn\fi\par\clearpage}

\def\section{\@startsection {section}{1}{\z@}{3.5ex plus 1ex minus
    .2ex}{2.3ex plus .2ex}{\normalsize\bf}}
\def\subsection#1{\subsectioncom{\sc{#1}}}
\def\subsectioncom{\@startsection{subsection}{2}{\z@}
    {3.25ex plus 1ex minus .2ex}{1.5ex plus .2ex}{\small}}
\def\subsubsection{\@startsection{subsubsection}{3}{\z@}{3.25ex plus
1ex minus .2ex}{1.5ex plus .2ex}{\small}}

\def\@addmarginpar{\@next\@marbox\@currlist{\@cons\@freelist\@marbox
    \@cons\@freelist\@currbox}\@latexbug\@tempcnta\@ne
    \if@twocolumn
        \if@firstcolumn \@tempcnta\m@ne \fi
    \else
      \if@mparswitch
         \ifodd\c@page \else\@tempcnta\m@ne \fi
      \fi
      \if@reversemargin \@tempcnta -\@tempcnta \fi
    \fi
    \ifnum\@tempcnta <\z@  \global\setbox\@marbox\box\@currbox \fi
    \@tempdima\@mparbottom \advance\@tempdima -\@pageht
       \advance\@tempdima\ht\@marbox \ifdim\@tempdima >\z@
      \else\@tempdima\z@ \fi
    \global\@mparbottom\@pageht \global\advance\@mparbottom\@tempdima
       \global\advance\@mparbottom\dp\@marbox
       \global\advance\@mparbottom\marginparpush
    \advance\@tempdima -\ht\@marbox
    \global\ht\@marbox\z@ \global\dp\@marbox\z@
    \vskip -\@pagedp \vskip\@tempdima\nointerlineskip
    \hbox to\columnwidth
      {\ifnum \@tempcnta >\z@
          \hskip\columnwidth \hskip\marginparsep
        \else \hskip -\marginparsep \hskip -\marginparwidth \fi
       \box\@marbox \hss}
    \vskip -\@tempdima
    \nointerlineskip
    \hbox{\vrule \@height\z@ \@width\z@ \@depth\@pagedp}}

\def\ref#1{
    \@ifundefined{r@#1}{{#1}\@warning{Reference `#1'
    on page \thepage \space
    undefined}}{\edef\@tempa{\@nameuse{r@#1}}\expandafter
    \@car\@tempa \@nil\null}}
\def\refn#1{\@ifundefined{r@#1}{{#1}\@warning{Reference `#1'
    on page \thepage \space
    undefined}}{\edef\@tempa{\@nameuse{r@#1}}\expandafter
    \@car\@tempa \@nil\null}}

\def\endequationl{\eqno \@eqnnum 
$$\global\@ignoretrue}

\def\eqnarray{\stepcounter{equation}\let\@currentlabel=\theequation
\global\@eqnswtrue
\global\@eqcnt\z@\tabskip\@centering\let\\=\@eqncr
$$\arraycolsep\z@
\halign to \displaywidth\bgroup\@eqnsel\hskip\@centering
  $\displaystyle\tabskip\z@{##}$&\global\@eqcnt\@ne
  \hskip 2\arraycolsep \hfil$\displaystyle{{}##{}}$\hfil
  &\global\@eqcnt\tw@ \hskip 2\arraycolsep
  $\displaystyle\tabskip\z@{##}$\hfil
   \tabskip\@centering&\llap{##}\tabskip\z@\cr}
\def\mmodetrue{\mmode=\iftrue}

\def\eqnarrayl#1{\stepcounter{equation}\let\@currentlabel=\theequation
\label {#1}
\global\@eqnswtrue
\global\@eqcnt\z@\tabskip\@centering\let\\=\@eqncr
$$\arraycolsep\z@
\halign to \displaywidth\bgroup\@eqnsel\hskip\@centering
  $\displaystyle\tabskip\z@{##}$&\global\@eqcnt\@ne
  \hskip 2\arraycolsep \hfil$\displaystyle{{}##{}}$\hfil
  &\global\@eqcnt\tw@ \hskip 2\arraycolsep
  $\displaystyle\tabskip\z@{##}$\hfil
   \tabskip\@centering&\llap{##}\tabskip\z@\cr}
\def\label#1{
\@bsphack\if@filesw {
{\ifnoncomplete{\makebox[1in][r]{\footnotesize{\sl [#1]}}}\fi}%
\let\thepage\relax
   \xdef\@gtempa{\write\@auxout{\string
      \newlabel{#1}{{\@currentlabel}{\thepage}}}}
}\@gtempa
   \if@nobreak \ifvmode\nobreak\fi\fi\fi\@esphack}

\def\newlabel#1#2{
\@ifundefined{r@#1}{}{\@warning{Label `#1' multiply
   defined}}\global\@namedef{r@#1}{#2}}

\def\endeqnarrayl{\@@eqncr\egroup
      \global\advance\c@equation\m@ne$$\global\@ignoretrue}

\newif\if@numbersec \@numbersectrue
\def\appendix{\par\clearpage
  \setcounter{section}{0}
  \setcounter{subsection}{0}
  \def\thesection{\Alph{section}}
  \def\thesubsection{\arabic{subsection}}
  \@ifstar{\def\@sectname{Appendix}\@numbersecfalse}
          {\def\@sectname{Appendix~}\@numbersectrue}}

\def\thefigures#1{\par\clearpage\section*{Figures\@mkboth
  {FIGURES}{FIGURES}}\list
  {Fig.~\arabic{enumi}.}{\labelwidth\parindent\advance\labelwidth -\labelsep
      \leftmargin\parindent\usecounter{enumi}}}

\def\thetables#1{\par\clearpage\section*{Tables\@mkboth
  {TABLES}{TABLES}}\list
  {Table~\arabic{enumi}.}{\labelwidth-\labelsep
      \leftmargin0pt\usecounter{enumi}}}

\def\@sect#1#2#3#4#5#6[#7]#8{\ifnum #2>\c@secnumdepth
     \def\@svsec{}\else
     \refstepcounter{#1}\edef\@svsec{\ifnum #2=1 \@sectname
         \if@numbersec\csname the#1\endcsname\fi.\else
         \csname the#1\endcsname.\fi
        \hskip 1em }\fi
     \@tempskipa #5\relax
      \ifdim \@tempskipa>\z@
        \begingroup #6\relax
          \@hangfrom{\hskip #3\relax\@svsec}{\interlinepenalty \@M #8\par}
        \endgroup
       \csname #1mark\endcsname{#7}\addcontentsline
         {toc}{#1}{\ifnum #2>\c@secnumdepth \else
                      \protect\numberline{\csname the#1\endcsname}\fi
                    #7}\else
        \def\@svsechd{#6\hskip #3\@svsec #8\csname #1mark\endcsname
                      {#7}\addcontentsline
                           {toc}{#1}{\ifnum #2>\c@secnumdepth \else
                             \protect\numberline{\csname the#1\endcsname}\fi
                       #7}}\fi
     \@xsect{#5}}

\def\@sectname{}

 \def\thefootnote{\fnsymbol{footnote}}

\def \@magscale#1{ scaled \magstep #1}
\font\frtnsfb = cmssbx10 \@magscale2 
\newcount\ieq
\newcount\jeq
\newcount\keq
\def \eq{
\multiply\ieq by 2
\jeq=\ieq
\divide\jeq by 4
\multiply\jeq by 4
\ifnum\ieq=\jeq \end{eqnarray} \keq=1 
\else
\keq=2 \begin{eqnarray} \fi
\ieq=\keq
}

\def \mathbox(#1){\invisible\ifmmode{{#1}}\else{\mbox{${#1}$}}\fi}
\def \mbf(#1){\mbox{\boldmath{$#1$}}}
\def \abs(#1){\mathbox(\left|{#1}\right|)}
\def \bracket(#1){\mathbox(\left\langle{#1}\right\rangle)}
\def \brav(#1){\mathbox(\langle {#1}|)}
\def \cg(#1,#2,#3,#4,#5,#6){\mathbox({(#1\,#2\,#3\,#4|#5\,#6)})}
\def \comm(#1,#2){\mathbox(\left[{#1},{#2}\right])}
\def \dfdx(#1,#2){\mathbox(\frac{{\rm d}{#1}}{{\rm d}{#2}})}
\def \delfdelx(#1,#2){\mathbox(\frac{\partial{#1}}{\partial{#2}})}
\def \inprod(#1,#2){\mathbox({(#1\cdot #2)})}
\def \inprodij(#1){\mathbox({\inprod(#1_i,#1_j)})}
\def \intd(#1,#2){\mathbox({\int^#1_#2 \; \rmd})}
\def \eps(#1){\mathbox(\epsilon_{#1})}
\def \half(#1){\mathbox(\frac{#1}{2})}
\def\onehalf{\half(1)}
\def \ketv(#1){\mathbox(|{#1}\rangle)}
\def \matele(#1,#2,#3){\mathbox(\left\langle {#1}|\,{#2}\,|{#3}\right\rangle)}
\def \mateled(#1,#2,#3){\mathbox(\left\langle
{#1}||\,{#2}\,||{#3}\right\rangle)}
\def \hatmbf(#1){\mathbox({\hat{\mbf({#1})}})}
\def \ninej(#1,#2,#3,#4,#5,#6,#7,#8,#9){\mathbox(\left\{\matrix
     {#1&#2&#3\cr#4&#5&#6\cr#7&#8&#9\cr}\right\})}
\def \rtov(#1,#2){\mathbox(\sqrt{{#1\over #2}})}
\def \sixj(#1,#2,#3,#4,#5,#6){\mathbox(\left\{\matrix
     {#1&#2&#3\cr#4&#5&#6\cr}\right\})}
\def \third(#1){\mathbox(\frac{#1}{3})}
\def \Trace(#1){\mathbox({\hbox{Tr} \left\{#1\right\}})}
\def \outprod(#1,#2){\mathbox({(#1\times #2)})}

\def \etal{{\it et al.}}
\def \etc{{\it etc.}}

\def \heart{\mbox{$\heartsuit$}}
\def \ie{{\it i.e.}}
\def \invisible{\mbox{$\rule{0mm}{1mm}$}}

\def \Lam{\mbox{$\Lambda$}}

\def \rmd{{\rm d}}

\def \siml{\raisebox{0.3ex}{$<$}\hspace{-0.8em}\raisebox{-0.3em}{$\sim$}}

\makeatother

\begin{document}

\def\wave{\simeq}
\def\rtt{\sqrt{3}}
\def\Rv{{\vec R}}
\def\MO{M.~Oka} \def\KY{K.~Yazaki}

\final \def\heart{}

\date{TIT/HEP-240/NP\\
      hep-ph/9312216\\
      November, 1993}

\title{Weak $\Lam N\to NN$ Transition in the Direct Quark Mechanism%
\footnote{talk presented by M.~Oka at {\sl the JSPS-NSF Joint Seminar on
``{\it Hyperon Nucleon Interactions}''}, Maui, HI, October, 1993}}

\author{%
Takashi Inoue\thanks{e-mail: tinoue@phys.titech.ac.jp},
Sachiko Takeuchi$^{(a)}$\thanks{e-mail: sachiko@phys.titech.ac.jp},
and
Makoto Oka\thanks{e-mail: oka@phys.titech.ac.jp}\\
{\sl Department of Physics, Tokyo Institute of Technology}\\
{\sl Meguro, Tokyo 152, Japan}\\
and\\
$^{(a)}${\sl Department of Public Health and Environmental Science}\\
{\sl Tokyo Medical and Dental University}\\
{\sl Yushima, Bunkyo, Tokyo 113, Japan}}

\maketitle

\abstract {
The weak $\Lam N\to NN$ transition is studied in the valence
quark model approach.  The momentum transfer for this transition
is so large that the short-distance two baryon dynamics must be taken
into account.  The two baryon system is described in the quark cluster
model and the weak transition amplitude is calculated by evaluating
the matrix elements of the effective weak $\Delta S= 1$
hamiltonian.  The results indicate some qualitative differences when
compared with those in conventional meson-exchange calculations.
Especially, we conclude that contributions of the $\Delta I={3\over2}$
transition are significant and that the discrepancy in the $n-p$ ratio
between theory and experiment could be resolved by including the direct-quark
processes.}

\thispagestyle{empty}

\newpage

\section{Introduction}

The hyperon $\Lambda$ decays weakly into a nucleon and a pion in the
free space.  Two isospin modes, $p\pi^-$ and $ n\pi^0$, share 64\% and
36\% of the total decay.  If the decay goes through a $\Delta I=
{1\over 2}$ vertex, the $p\pi^-/n\pi^0$ ratio would be two to one except for
a small correction due to the phase space difference.
The experimental ratio is very close to the $\Delta I= {1\over 2}$
prediction, and thus support the $\Delta I= {1\over 2}$ hypothesis
for the hadronic weak decay.

In the nuclear medium, the $\Lambda\to N\pi$ decay is suppressed by
the Pauli blocking on the final nucleon state, whose momentum is
less than 100 MeV/c for the $\Lambda$ decay at rest.  Indeed, in heavy
hypernuclei, the decay is predominantly the nonmesonic one, that is,
$\Lambda N \to NN$.  If we assume that the initial $\Lambda$ and the
nucleon are at rest, then the final relative momentum of $NN$ is
about 420 MeV/c and thus is well above the Fermi momentum.

{\heart}The purpose of this report is to study the direct quark processes in
the two-body $\Lambda N\to NN$ weak decays and
to show qualitative differences
from the conventional picture employing the meson exchange mechanism.
We present a possibility to solve the problems that the
meson exchange mechanism encounters.

The nonmesonic decays of hypernuclei seem quite useful in studying the
low energy nonleptonic weak interactions among quarks.  The final
relative momentum of $NN$ is high enough to look into the
short-distance component of the two nucleon system.  This type of the
hyperon decay may reveal a new aspect of the weak interaction under
the influence of the strong interaction.
An advantage of using the hypernuclear decay
is that the process is selective in isospin, spin and
orbital angular momentum for appropriate initial and
final states  of the hypernucleus.

Theoretical study of the nonmesonic decay of hypernuclei has traditionally
employed the meson ($\pi$, $K$, $\rho$, etc.) exchange mechanism, where one
of the meson-baryon vertices involves the weak transition $s\to
d${\cite{meson}}.
Accumulating experimental data, however, have
revealed some difficulties in the meson-exchange picture.  For instance,
the so-called $n-p$ ratio, i.e., the ratio $R_{np}$ of $\Lambda n \to
nn$ v.s. $\Lambda p \to np$ decay in the nucleus, is predicted very
small,  $R_{np}\wave 0.1-0.4$ in the meson-exchange picture.  This is due
to the strong contribution of the tensor force, which is preferred at the
large momentum transfer.  The tensor force selects the $S=1$, $I=0$
$pn$ final state and therefore $R_{np}$ becomes small.
The experimental data seem not to agree with the prediction, i.e.,
$R^{exp}_{np} \wave 1$ in decays of light hypernuclei.
We argue that the direct quark process, which does not follow the
$I=0$ selection rule, may enhance the $n-p$ ratio.

The mesonic weak decays of hyperons have been tested for the $\Delta
I= {1\over 2}$ rule and are known to satisfy the rule to about 5\%
error.   The same rule for the nonmesonic weak processes, like
$\Lambda N \to NN$, is not confirmed yet.
Indeed, an analysis of the
decay of the  $A=3$ and 4 hypernuclei claims that the $\Delta I=
{1\over 2}$ rule may be violated{\cite{Shoe}}.  It is therefore
urgent to clarify the mechanism of the
$\Delta I= {1\over 2}$ rule in the free hyperon decays and to study
whether the same mechanism restricts the nonmesonic decays to $\Delta
I= {1\over 2}$ as well.

In the study of the meson-exchange processes, the $\Delta I= {1\over
2}$ rule is assumed from the beginning, implemented in the $\Lambda\to
N\pi$ vertex.  We instead employ the effective quark-quark weak
hamiltonian, which contains both the $\Delta I= {1\over 2}$ and
$\Delta I= {3\over 2}$ components.  Although the $\Delta I=
{3\over 2}$ part has a small overall coefficient, we will see that its
matrix elements for the $\Lambda N\to NN$ decay may not be small
compared to the $\Delta I= {\onehalf}$ component.

\medskip
The paper is organized as follows.  Sect~2 is devoted to the
basic formulation of the present calculation.  The effective
hamiltonian and the quark cluster model wave functions are presented.
Various approximations employed in this calculation are examined.
We present the results of the calculation in sect.~3.  They are used
for studying qualitative differences between the direct-quark
mechanism and the conventional meson-exchange picture.
A conclusion is given in sect.~4.

\section{Formulation}
\subsection{Effective Weak Hamiltonian}

The effective weak hamiltonian describing $\Delta S = 1$ processes
has been calculated by several authors{\cite{VSZ,Okun,Paschos}}.
It can be computed by
analyzing the correction due to the strong interaction on the pure
weak vertex $su \to du$:
\eq
   H(\hbox{purely weak:}\Delta S =1) =
        - {G_f\over \sqrt{2}}\, \sin \theta_c \,
           (\bar u_{\alpha}s_{\alpha})_{V-A}
           (\bar d_{\beta}u_{\beta})_{V-A}
\label{eq:weak}
\eq
where
\eq
(\bar u_{\alpha}s_{\alpha})_{V-A} \equiv (\bar u_{\alpha}
        \gamma^{\mu}(1-\gamma_5) \, s_{\alpha}) \quad\rm{\etc}
\eq
and $\alpha$ and $\beta$ denote the color of quarks and the
color sum is always assumed.
Note that there exists no strangeness-changing neutral current in the
standard electro-weak theory, and thus the vertex is only for the
left-handed quarks.

This purely weak four-quark vertex clearly contains both the $\Delta I
= \onehalf$ and $\Delta I= {3\over 2}$ components.
It was pointed out, however, that the correction due to the strong
interaction enhances the $\Delta I = \onehalf$ component while  the $\Delta
I= {3\over 2}$ is suppressed at the same time{\cite{VSZ}}.
The strong correction is treated
perturbatively at the scale $Q^2 \wave M_W^2$, and only the lowest
order diagrams are taken into account.
The mechanism of the $\Delta I= \onehalf$ enhancement can be
understood by realizing the anomalous dimensions of the
$\Delta I= \onehalf$ and $\Delta I= {3\over 2}$ components in
eq.(\ref{eq:weak})  have opposite
signs with each other.  The $\Delta I= \onehalf$ ($\Delta I= {3\over 2}$)
anomalous dimension is positive (negative)
and therefore, when the
renormalization scale is moved down from the $W$ boson mass to the
low-energy hadronic scale ($\siml 1$ GeV), the operator with the positive
anomalous dimension is enhanced and vice versa.
{\heart}The renormalization group equation also induces new four-quark
operators through operator mixings.
Another contribution comes from the so-called penguin diagrams.
They are purely $\Delta I= \onehalf$ and thus help the $\Delta I=
\onehalf$ rule.

{\heart}Taking these effects into account, the low energy
effective weak hamiltonian has been derived{\cite{Paschos}}:
\eq
  H_{eff}^{\Delta S=1}\left(Q^2\sim{\mu}^2\right)=
  -\frac{G_f}{\sqrt 2}\sum_{r=1,r\ne 4}^6K_rO_r
\eq
where the four-quark operators, $O_k$ ($k= 1$, 2, 3, 5 and 6) are
defined by{\cite{VSZ}}{\heart}
\eq
 O_1 &=& (\bar d_{\alpha}s_{\alpha})_{V-A}
          (\bar u_{\beta}u_{\beta})_{V-A}
         -(\bar u_{\alpha}s_{\alpha})_{V-A}
           (\bar d_{\beta}u_{\beta})_{V-A}\\
  O_2 &=& (\bar d_{\alpha}s_{\alpha})_{V-A}
           (\bar u_{\beta}u_{\beta})_{V-A}
         +(\bar u_{\alpha}s_{\alpha})_{V-A}
           (\bar d_{\beta}u_{\beta})_{V-A}\nonumber\\
     &+& 2(\bar d_{\alpha}s_{\alpha})_{V-A}
         (\bar d_{\beta}d_{\beta})_{V-A}
        +2(\bar d_{\alpha}s_{\alpha})_{V-A}
         (\bar s_{\beta}s_{\beta})_{V-A}\\
  O_3
     &=& O_3(\Delta I={1\over 2}) + O_3(\Delta I={3\over 2}) \\
  O_3(&\Delta& I={1\over 2}) = {1\over 3}\left[
           (\bar d_{\alpha}s_{\alpha})_{V-A}
           (\bar u_{\beta}u_{\beta})_{V-A}
         + (\bar u_{\alpha}s_{\alpha})_{V-A}
           (\bar d_{\beta}u_{\beta})_{V-A}\right.\nonumber\\
     &+& \left.2 (\bar d_{\alpha}s_{\alpha})_{V-A}
         (\bar d_{\beta}d_{\beta})_{V-A}
        -3 (\bar d_{\alpha}s_{\alpha})_{V-A}
         (\bar s_{\beta}s_{\beta})_{V-A}\right]\nonumber\\
  O_3(&\Delta& I={3\over 2}) = {5\over 3}\left[
           (\bar d_{\alpha}s_{\alpha})_{V-A}
           (\bar u_{\beta}u_{\beta})_{V-A}
         + (\bar u_{\alpha}s_{\alpha})_{V-A}
           (\bar d_{\beta}u_{\beta})_{V-A}\right.\nonumber\\
     &-& \left. (\bar d_{\alpha}s_{\alpha})_{V-A}
         (\bar d_{\beta}d_{\beta})_{V-A}\right]\nonumber\\
  O_5 &=& (\bar d_{\alpha}s_{\alpha})_{V-A}
       (\bar u_{\beta}u_{\beta}+\bar d_{\beta}d_{\beta}
      + \bar s_{\beta}s_{\beta})_{V+A}\\
  O_6 &=& (\bar d_{\alpha}s_{\beta})_{V-A}
       (\bar u_{\beta}u_{\alpha}+\bar d_{\beta}d_{\alpha}
      + \bar s_{\beta}s_{\alpha})_{V+A}
\eq
Among these operators,  $O_3$ contains a part that induces the $\Delta
I={3\over 2}$ transition, $O_3(\Delta I={3\over 2})$, while the others
are purely $\Delta I=\onehalf$.
{\heart}The coefficients $K_r$ can be
calculated by solving the renormalization group equation to
the one-loop QCD corrections.
They depend on the mass of the top
quark $m_t$ through the penguin diagrams, which generate the
operators $O_5$ and $O_6$ with the $V+A$ coupling.
We find that the final results are insensitive to the choice
of $m_t$  and  here choose $m_t= 200 $ GeV/c$^2$.
The coefficients also depend on
the energy scale $\mu^2$ for the effective
hamiltonian.
In the present calculation, we choose two sets of the values given in
ref.\cite{Paschos}: $\mu=0.24$ GeV and $0.71$ GeV.
They are chosen so as to give $\alpha_s (\mu^2)=1 $ for the QCD
$\Lambda_{QCD} $ parameter,
$\Lambda_{QCD} = 0.1$ GeV and 0.316 GeV, respectively.
The values of the coefficients $K_r$
used in the present calculation are given in Table 1.
One sees that the two choices are not much different
except for $K_5$ and $K_6$.
We will find that the differences in the transition
matrix elements for these choices are at most 10\%.

\begin{table}
\caption{
Two choices of strengths of the weak effective four-fermi vertices,
taken from ref.[5].  We use the version with flavor-number
dependent $\Lambda$ and $m_t=200$GeV. The values of the CKM matrix
elements are taken as the central values of those given in ref.[6].}
\begin{center}
\begin{tabular}{ccc|ccccc}
 \hline
 &$\mu $ (GeV) & $\Lambda^{(4)} $ (GeV) & $K_1$ & $K_2$ & $K_3$ &  $K_5$  &
 $K_6$\\
 \hline
 I&0.24& 0.10& $-0.284$ & $0.009$  & 0.026 & 0.004& $-0.021$\\
 II&0.71& 0.316& $-0.270$ & $0.011$  & 0.027 & 0.002& $-0.010$\\
 \hline
\end{tabular}
\end{center}
\end{table}

The most prominent feature of this effective hamiltonian is that the
QCD correction enhances the $O_1$ component while the other terms are
suppressed. This is the main mechanism for the $\Delta I= \onehalf$ enhancement
as is explained above.
Later we will compare the results with and without $O_3(\Delta I= {3\over 2})$
in order to study the $\Delta I= {3\over 2}$
contribution.

This effective hamiltonian has been used for the calculations of the
nonleptonic decay of strange mesons and baryons{\cite{Okun}}.
It is found that although the $\Delta I= \onehalf$ transition is
indeed enhanced in those  decays,  the enhancement is not enough to
account for the experimental data quantitatively.
It was suggested{\cite{Sanda}} that an additional $\Delta I=\onehalf$
enhancement  arises from the mesonic correction in the
chiral effective theory.
It was also suggested that the decay
amplitudes may be sensitive to the meson and
baryon wave functions{\cite{Fujii}}.

\subsection{Six-Quark Wave Function}
In calculating the decay amplitude for $\Lambda N \to NN$, we employ
the constituent quark model, which describes the spin-flavor structure
of the ground-state baryons very well.
Two-baryon systems are expressed by
the quark-cluster-model wave functions{\cite{OY80,Theo1}.
First, we assume that the baryon consists of
three valence quarks, whose orbital wave function is a harmonic
oscillator eigenstate,
\eq
\phi(1,2,3)^{\mbox{orb}}=
\left(\frac{1}{2\pi b^2}\right)^{\frac34}
   \left(\frac{2}{3\pi b^2}\right)^{\frac34}
    \mbox{exp}\left\{-\frac{1}{4b^2}{\vec{\xi}_{12}}^2\right\}
   \mbox{exp}\left\{-\frac{1}{3b^2}{\vec{\xi}_{12-3}}^2\right\}
\eq
where $\xi$'s are the Jacobi coodinates and the Gaussian
parameter $b$ is chosen $b=0.5$ fm.
{\heart}We here neglect the asymmetry due to the
mass difference of the s and u quarks in the \Lam\ wave function.
The six quark wave functions are given by
\eq
\vert \Lambda N\rangle
                      &= &{\cal A}^6\vert \phi(1,2,3)\phi(4,5,6)
                          \chi_0(\vec{R})\rangle\nonumber\\
\vert NN\rangle
               &=&{\cal A}^6\vert \phi(1,2,3)\phi(4,5,6)
               \chi(\vec{R})\rangle
\label{eq:WF}
\eq
where ${\cal A}^6$ is the antisymmetrization operator for six quarks,
$\phi$ is the internal wave function of the baryon, and
$\vec{R}$ is the relative coodinate of two baryons.
$\chi_0(\Rv)$ ($\chi(\Rv)$)  is the initial (final)
relative wave function.
The flavor-spin part of $\phi$ is taken to be purely the SU(6)
wave function.

In the present calculation,
we choose the simplest relative wave functions, \ie,
a Gaussian for the initial state  and the plane wave for the final state.
\eq
\chi_0(\vec{R})&=&\left(\frac{1}{\pi B^2}\right)^{\frac34}
          \exp \left\{-\frac{1}{2B^2}\vec{R}^2\right\} \\
\chi(\vec{R}) &=&
          \exp \left\{i\vec{k}\cdot\vec{R}\right\}
\eq
The relative momentum of the final state is determined by the
realistic $Q$ value $\Delta E\equiv M_{\Lambda} - M_N$ of the decay:
$k=416$ MeV/c.
The Gaussian $B$ parameter of the initial state is to be determined by
the $\Lambda N$ wave function in the hypernucleus.  Suppose that we choose
(unrealistically) $B= \sqrt{2/3} \,b$, then the initial wave function is
reduced to the $(0s)^6$ configuration in the harmonic
oscillator shell model.  This can be interpreted as a dibaryon state
in which the initial $\Lambda$ and $N$
are on top of each other ($R=0$).
Later we show the results of the calculation where we employ  $B= 1.838$
fm ($=\sqrt{2}\times$ 1.3), that corresponds to a $\Lambda N$ system in
the $^4$He nucleus.

Although these choices of the wave functions may not be totally realistic,
they will clarify the qualitative difference between
the meson-exchange and the direct-quark processes, which is the
purpose of this study.  An advanced
calculation using the realistic two-baryon wave functions is under way.

\medskip
Because we employ the nonrelativistic valence quark picture for the
wave functions, the effective hamiltonian is also approximated by
adopting the Breit-Fermi nonrelativistic expansion up to $p/m$.
Then the spin-orbital part of the operator contains terms which
conserve parity,
\eq
1\times 1,\quad
\left(\vec{\sigma}_i\cdot\vec{\sigma}_j\right)\times 1
\nonumber
\eq
and those which break parity,
\eq
    \left(\vec{\sigma}_i\pm\vec{\sigma}_j\right)\cdot\vec{q}_{ij},\quad
      \left(\vec{\sigma}_i\pm\vec{\sigma}_j\right)
             \cdot\left(\vec{P}_i\pm\vec{P}_j\right),\quad
   i\left(\vec{\sigma}_i\times\vec{\sigma}_j\right)
                   \cdot\vec{q}_{ij} ,\quad
     i\left(\vec{\sigma}_i\times\vec{\sigma}_j\right)  \cdot
                   \left(\vec{P}_i\pm\vec{P}_j\right)
\nonumber
\eq
where $\vec q_{ij} = \vec p'_i - \vec p_i$ and $\vec P_i= \onehalf (\vec
p_i + \vec p'_i)$ with $\vec p_i$ ($\vec p'_i$) the initial (final)
momentum of the $i$-th quark.
{\heart\heart}In this expansion, we take account of the SU(3) breaking
effects due to the quark mass differences.
Then the coefficients for these
operators depend on the light quark constituent mass, $m_q=m_u=313$ MeV
and the mass ratio, $m_u/m_s =0.6$.

\begin{table}
\caption{Possible initial and final quantum numbers
for the initial $L=0$ transition}
\begin{center}
\begin{tabular}{cccccc}
\hline
\mbox{channel}&  \mbox{isospin}  &  \mbox{spin--orbital} \\
\hline
1 &   $p\Lambda\to pn$   &   $^{1}S_0\to {}^{1}S_0$  & $a_p$ \\
2 &                      &   $^{1}S_0\to {}^{3}P_0$  & $b_p$ \\
3 &                      &   $^{3}S_1\to {}^{3}S_1$  & $c_p$ \\
4 &                      &   $^{3}S_1\to {}^{3}D_1$  & $d_p$ \\
5 &                      &   $^{3}S_1\to {}^{1}P_1$  & $e_p$ \\
6 &                      &   $^{3}S_1\to {}^{3}P_1$  & $f_p$ \\
\hline
7 &   $n\Lambda\to nn$   &   $^{1}S_0\to {}^{1}S_0$ & $a_n$ \\
8 &                      &   $^{1}S_0\to {}^{3}P_0$ & $b_n$ \\
9 &                      &   $^{3}S_1\to {}^{3}P_1$ & $f_n$ \\
\hline
\end{tabular}
\end{center}
\end{table}

In the present study,  we restrict our initial state to $L=0$.
Table 2 shows nine possible
combinations of $L$, $S$, $J$, and $I$  for the
initial and final states.
We note that the $I=1$ final states are allowed both for
($\Lambda n \to nn$) and ($\Lambda p \to pn$), while the $I=0$
states are not possible for ($\Lambda n \to nn$).
Thus  we have
6 ($\Lambda p \to pn$) and 3 ($\Lambda n \to nn$) matrix elements,
which are labeled from $a$ through $f$ in Table 2, according to the
widely used notation{\cite{BD}}.
Among them, the channels
1, 3, 4, and 7 ($a$, $c$ and $d$) are the parity conserving
transitions, while the others violate the parity invariance.
The amplitude $d$ is not zero only for the tensor component of the weak
interaction, which we neglect in our quark model calculation by
truncating the $p/m$ expansion.

The  transition amplitudes are calculated
according to the standard quark cluster model approach.  Remember that
we take into account the full antisymmetrization among six quarks.
One needs to calculate the exchange matrix elements as well as the
direct ones.

\section{Results}

The results of the calculation are summarized in Table 3.
Nine amplitudes give all the information for the $\Lambda N\to NN$ weak
decay from $L=0$.
We compare the results for the two choices of the weak hamiltonian parameters,
which differ in the energy scale or $\Lambda_{QCD}$, given in Table 1 and
see that their differences are small.
The numbers
given under the ``$\Delta I= {3\over 2}$ omitted'' are the results
without  the
$\Delta I= {3\over 2}$ component of the $O_3$ operator.
In this case, the ratio $a_n/a_p$, $b_n/b_p$, and $f_n/f_p$
are equal to $\sqrt 2$.
The other amplitudes, $c$, $d$ and $e$ do not contain any $\Delta I=
{3\over 2}$ component, because the final $NN$ states have $I=0$.
We find that the amplitudes $a$ and $b$ get significant
contributions from the $\Delta I= {3\over 2}$ component, while
$f$ has only a small contribution of $\Delta I= {3\over 2}$.
Thus we conclude that the $\Delta I= {3\over 2}$ transition can be
studied in the weak decay starting from the $\Lambda N$ $^1S_0$ state.

\begin{table}
\caption{Calculated transition matrix elements in $10^{-10}$ $MeV^{-1/2}$.
The numbers without parenthesis (in parenthesis) are the results for
the parameter I (II).}
\begin{center}
\begin{tabular}{c|rr|rr|r}
\hline
channel&\multicolumn{2}{c|}{full}&
       \multicolumn{2}{c|}{$\Delta I=\frac32$ omitted}& OPE \\
\noalign{\hrule}
$a_p$ & $-6.66$ &($-6.83$)  & $-0.02$ &($0.04$)    &  4.52  \\
$b_p$ & $5.79$ &($6.30$)     & $0.06$ &($0.37$)      &  $-24.4$  \\
$c_p$ & $2.70$ &($2.48$)     & $2.70$ &($2.48$)      &  $4.52$  \\
$d_p$ & $0$ &($0$)  & $0$ &($0$) &  $-83.0$  \\
$e_p$ & $-2.21$ &($-2.01$)  & $-2.21$ &($-2.01$)  &  $-42.3$  \\
$f_p$ & $-5.57$ &($-5.02$)  & $-5.39$ &($-4.83$)  &  $19.9$  \\
\noalign{\hrule}
$a_n$ & $4.67$ &($4.92$)     & $-0.03$ &($0.06$)   &  $6.39$  \\
$b_n$ & $-3.96$ &($-3.67$)  & $0.09$ &($0.52$)    &  $-34.5$  \\
$f_n$ & $-7.49$ &($-6.70$)  & $-7.62$ &($-6.83$) &  $28.2$  \\
\noalign{\hrule}
\end{tabular}
\end{center}
\end{table}

\begin{table}
\caption{Calculated observables.
The numbers without parenthesis (in parenthesis) are the results for
the parameter I (II).}
\begin{center}
\begin{tabular}{c|rr|rr|r}
\hline
  & \multicolumn{2}{c|}{full}& \multicolumn{2}{c|}{$\Delta I=\frac32$ omitted}
  & OPE\\
\noalign{\hrule}
$\Gamma_p$  ($10^{7}$ {sec}$^{-1})$&  0.39  &(0.36) &  0.23 &(0.19) & 52.4 \\
$\Gamma_n$  ($10^{7}$ {sec}$^{-1})$&  0.39  &(0.32) &  0.33 &(0.26) & 6.8 \\
\noalign{\hrule}
$R_{np}$        & 0.99 &(0.89)         & 1.41 &(1.39)
                & 0.13 \\
$\eta_p$        & 2.13 &(1.96)         & 4.64  &(4.47)
                & 0.34 \\
$\eta_n$        & 8.43 &(6.12)         & 2.23$\cdot 10^5$&(4.04$\cdot10^4$)
                & 87.5 \\
\noalign{\hrule}
$a_1(p)$        & $-0.36$ &($-0.32$)   & $-0.58$ &($-0.58$)
                & $-0.19$\\
\noalign{\hrule}
\end{tabular}
\end{center}
\end{table}

In Table 4, we summarize the calculated decay rates with the initial
spin averaged and the final states summed up.
\eq
   \Gamma_p &=&\frac{\pi M_N k}{(2\pi)^3}\,
           {1\over4} \, [a_p^2 + b_p^2 + 3(c_p^2 + d_p^2 + e_p^2 +f_p^2)]  \\
   \Gamma_n &=&\frac{\pi M_N k}{(2\pi)^3}\,
           {1\over4} \, [a_n^2 + b_n^2 + 3f_n^2 ]
\eq
We again find that the $\Delta I= {3\over 2}$ component of $O_3$
changes the total decay rate by as much as a factor two
for the direct quark processes.

The $n-p$ ratio,
\eq
R_{np} \equiv \Gamma_n/\Gamma_p ,
\eq
the ratio of the parity
violating (PV)  v.s. the parity conserving (PC) contributions,
\eq
    \eta_p&=& {b_p^2+3(f_p^2+e_p^2) \over a_p^2 +3(c_p^2+d_p^2)}  \\
    \eta_n&=& {b_n^2+3f_n^2\over a_n^2 }
\eq
and the decay asymmetry parameter, $a_1$,
are also given in Table 4.
We find that the $n-p$ ratio for the direct quark process is much
larger than that obtained in the meson-exchange calculation.
{\heart}The pure $\Delta I={1\over 2}$ calculation also yields large $R_{np}$,
indicating that the $R_{np}$ enhancement is not related to the $\Delta
I=\onehalf$ rule violation.

The results of the one-pion exchange transition are also shown in
Tables 3 and 4, for comparison{\cite{meson}}.
These amplitudes satisfy
$a_n/a_p=b_n/b_p=f_n/f_p=\sqrt2$,
because $\Delta I=\onehalf$ is assumed for the weak pion-baryon vertex.
One sees that the amplitude $d$ in this case is dominant.  This comes from
the tensor part of the one pion exchange and is enhanced due to a
large relative momentum in the final state.
Because the amplitude $d$ is not allowed for the $n\Lambda \to nn$ by
the Pauli principle,  the $n-p$ ratio in the pion exchange amplitudes
becomes very small.
Remember that the tensor
part of the direct quark interaction is neglected because it is of order
$(p/m)^2$.

The magnitudes of the transition amplitudes are in general larger for
the meson exchange mechanism than the direct-quark process.  It is
noticed, however, that some amplitudes, such as $a$ and $c$, have
comparable direct-quark amplitudes.
Therefore, if one can select the initial and/or final spin
states in the decay experiments,  it will be possible to detect
the contribution of the direct quark processes.

The decay asymmetry parameter describes the angular distribution of
the outgoing two nucleons in the rest frame,
\eq
	W(\theta) = 1 + a_1 \, {\cal P}_{\Lambda} \, P_1(\cos\theta)
\eq
where ${\cal P}_{\Lambda}$ is the polarization of \Lam\
in the nucleus.  The parameter $a_1$ is given in terms of the two-body
decay amplitudes by
\eq
      a_1 =  {2\sqrt{3}\,(\sqrt{2} c+d) f\over a^2+b^2+3(c^2+d^2+e^2+f^2) }
\eq
and thus indicates the interference between the PV and PC components.
For the $\Lambda n \to nn$ decay, $c$ and $d$ vanish due to the isospin
conservation and therefore $a_1(n)$ is zero.
{\heart}Recent experiment done at KEK indicates a large negative
$a_1(p)$ for light hypernuclei\cite{Ajimura}.  The data is consistent
with
$a_1(p) \le -0.6$.
Our calculation yields the correct sign, but the magnitude is
smaller.  The meson exchange calculation done by Ramos \etal, also
predicted a small magnitude with the correct sign{\cite{meson}}.

{\heart}Recently, Schumacher{\cite{Shoe}} suggested that analyses
of the non-mesonic decays of the A=4 and 5 hypernuclei may be usuful
in checking the $\Delta I=\onehalf$ rule for the nonmesonic weak decay.
One can parametrize the n-p ratios of the
decay of ${}^4_{\Lambda}$He and ${}^5_{\Lambda}$He
and the ratio of non-mesonic decay widths of
${}^4_{\Lambda}$He and ${}^4_{\Lambda}$H in terms of
$R_{nS}$ and $R_{pS}$, where $R_{NS}$ stands for the decay rate of
$\Lambda N$ with the initial spin $S${\cite{BD}}.
Then, using experimental data, the ratios of
$R_{NS}$ can be extracted.
In fact, the $R_{n0}/R_{p0}$ ratio will be 2 for the pure
$\Delta I=\onehalf$ transition, while it becomes 1/2 for the pure
$\Delta I=3/2$ transition.
{\heart}Because these hypernuclei involve only the relative s-wave
states, the direct quark processes can play significant roles.
It may then be possible to see the $\Delta I=3/2$ decay, if any.
Table 5 shows the $J=0$ ($S=0$) part of the n-p ratio,
$R_{n0}/R_{p0}$ as well as the above observables for our direct quark
amplitudes with and without the $\Delta I={3\over 2}$ components.
{\heart}Our full amplitudes indeed predict a small ratio $R_{n0}/R_{p0}$,
which indicated a large $\Delta I=3/2$ contribution.
It is therefore concluded that the $\Delta I={3\over 2}$ components of
the direct quark processes
is possibly observed in non-mesonic decays of light hypernuclei.

\begin{table}
\caption{Observables for light hypernuclei.
The numbers without parenthesis (in parenthesis) are the results for
the parameter I (II).}
\begin{center}
\begin{tabular}{r|rr|rr|r}
 & \multicolumn{2}{c|}{full}&\multicolumn{2}{c|}{ $\Delta I=\frac32$ omitted}
 & OPE \\
\noalign{\hrule}
$R_{np}({}^4_{\Lambda}$He)&  0.36  &(0.39) & 0 &(0) & 0.09  \\
$R_{np}({}^5_{\Lambda}$He)&  0.99  &(0.89) & 1.41 &(1.40) & 0.13 \\
$      {\Gamma_{n.m.}(^4_{\Lambda}\mbox{He})}/
       {\Gamma_{n.m.}(^4_{\Lambda}\mbox{H} )}$
   & 0.78  &(0.78) & 0.71 &(0.72)    & 6.26  \\
\noalign{\hrule}
${R_{n0}}/ {R_{p0}}$ & $0.48$ &($0.44$) & $2$ &($2$) & $2$ \\
\noalign{\hrule}
\end{tabular}
\end{center}
\end{table}

\section{Conclusion and Discussion}

We present a  quark model calculation of the direct-quark
processes of the weak $\Lambda N\to NN$ decay, which can be observed
exclusively in decays of hypernuclei.
{\heart}We find that the transition amplitudes in some of the
decay channels are comparable to those in the meson exchange decays in
magnitudes
and show qualitatively distinctive properties.
This is encouraging because the direct-quark processes may resolve the
discrepancies between experiment and theory based on the
meson-exchange mechanism.
Indeed, after averaging over the initial and final spin
states, we obtain a large $n-p$ ratio from the direct-quark
amplitudes.
In the actual hypernuclear decays, one expects a variety of the
spin-isospin combinations.
We also have to note that the initial
orbital angular momentum greater than zero may contribute
significantly.  We, therefore, cannot make a definite prediction here.
Further study of the $\Lambda  N\to NN$ decay with higher partial waves
and realistic two-baryon wave functions is under way.

It is also important to combine the meson-exchange amplitudes
with the direct-quark ones to make a final quantitative conclusion.
We here do not superpose the direct-quark and the one-pion-exchange
amplitudes because both of them are yet incomplete.
The one-pion exchange is far from realistic because other mesons, such
as $K$ and $\rho$, are known to contribute to this process significantly.
{\heart}It is not clear whether the phenomenological lagrangian
for the mesonic decay of $\Lambda$ is consistent with the
weak effective  hamiltonian used in the direct-quark calculation.
It is important to understand the mesonic decay and its
$\Delta I=\onehalf$ rule starting from the effective four-quark
Lagrangian{\cite{Okun,Sanda}}.
A study along this line is being carried out.

Our plan also includes the study of $\Lambda N \to NN$ with more realistic wave
functions for the initial $\Lambda N$ and the final $NN$ states.
Especially, effects of the baryon-baryon short-range correlation may
change the results quantitatively.
A more realistic calculation is under way.

\end{document}